\documentclass[%
 reprint,
showpacs,
 amsmath,amssymb,
 aps,
pre,
]{revtex4-1}
 
\usepackage{xcolor}
\usepackage{comment}

\usepackage{graphicx}
\usepackage{dcolumn}
\usepackage{bm}

\begin{document}

\preprint{APS/123-QED}

\title{Evolution of instabilities in filament buckling processes}

\author{A. G. Monastra$^{1,2}$}
\email{amonast@ungs.edu.ar}

\author{M. F. Carusela$^{1,2}$}
\author{G. van der Velde$^{3}$}
\author{M. V. D'Angelo$^{4,2}$}
\author{L. Bruno$^{3,2}$}

\affiliation{$^1$ Instituto de Ciencias, Universidad Nacional de Gral. Sarmiento, Los Polvorines, Buenos Aires, Argentina}
\affiliation{$^2$ Consejo Nacional de Investigaciones Cient\'\i ficas y T\'ecnicas, Argentina}
\affiliation{$^3$ Departamento de F\'\i sica \& IFIBA-CONICET, Facultad de Ciencias Exactas y Naturales, Universidad Nacional de Buenos Aires, Argentina}
\affiliation{$^4$ Grupo de Medios Porosos, Facultad de Ingenier\'\i a, Universidad Nacional de Buenos Aires, Argentina\\}

\date{\today}

\begin{abstract}

In this work we study the dynamical buckling process of a thin filament immersed in a high viscous medium.
We perform an experimental study to track the shape evolution of the filament during a constant velocity compression.  Numerical simulations reproduce the dynamical features observed for the experimental data and allow quantifying the filament's load.

We observe that both the filament's load and
the wavenumber evolve in a step-wise manner. To achieve a physical insight of the process we apply a theoretical model to describe the buckling of a filament in a viscous medium. We solve a hydrodynamic equation in terms of normal modes for clamped-clamped boundary conditions and applied constant load. We find a good agreement with experimental data and simulations, suggesting that the proposed mechanistic model captures the essential features underlying the dynamical buckling process.

 \end{abstract}








\maketitle


\section{Introduction}

The buckling of filaments under compression in high viscous media  is a phenomenon of interest in numerous applications in physics, biology, medicine and engineering, and involve a wide range of spatial and temporal scales. Examples include the manufacturing of fiber-reinforced composites \cite{Yasuda, DooYeo}, the rheology of biological polymers \cite{Lagomarsino},  the motility of microorganisms \cite{Lowe, Chaban} and the movement of microtubules of the cytoskeleton \cite{Brangwynne}. Polymeric solutions allow to improve the displacement in enhanced oil recovery \cite{Xie} and fibers are used to prevent proppant flow back in hydraulic fracturing \cite{Liang} in the oil industry, and optical fibers are employed as sensors for measurements in groundwater \cite{Selker}. In all these processes, the movement of the flexible filaments in a confined environment can produce their entrapment or buckling.

The buckling process is an instability that occurs when a compressing force is applied to the end of a slender rod. If the applied force exceeds a critical value \cite{LandaLifschitz, Timoshenko} the originally straight filament collapses and buckles. The Euler or critical force depends on the rod geometry and its elasticity. For a clamped-clamped filament its value is:

\begin{equation}
P_c = 4 \pi^2 \frac{EI}{L^2} \ ,
\end{equation}
where $E$, $I$ and $L$ are the filament's Young modulus, the second moment of inertia and length, respectively. 

Whereas the critical force does not depend on the environment where the buckling takes place, the time scale of the subsequent filament deformation strongly depends on the surrounding viscosity \cite{Howard2001}. Thereby, the time course of the shape evolution will be slower for a very viscous fluid.  

As mentioned before the motion of filaments in viscous media is present in several natural phenomena and industrial applications. Typically in these cases the applied load is not constant and thus an equilibrium situation is not attained. For this reason it is of great interest to understand the conditions under which a flexible filament undergoes deformation, as well as its characteristic shape and amplitude. Such deformations could be responsible, for example, of the entrapment of the filament, preventing its transport through the medium.

In this work we register the buckling of a thin rod when it is compressed  in a very viscous medium, i.e. glycerol, at constant speed. We recover the filament shapes from the movies and compute the wavenumber as the deformation evolves for different compressing speeds.  
To relate these shapes with the load, we develop a numerical simulation of the process and obtain a relationship between the wavenumber and the load.
To achieve a physical understanding, we also develop a mechanical theoretical model of a filament under a constant load. This model is based on normal modes with characteristic wavenumbers that are directly related to the compressing force.

\section{Mechanical model}

The configuration of a uniform slender semi-flexible filament is determined by the position of its neutral axis ${\bf r} (l)$, with $l$ a curvilinear coordinate along the filament varying from 0 to $L$ (unstressed length). The configuration has elastic and bending energies given by:

\begin{eqnarray}
V_{\mathrm{E}} &=& \frac{1}{2} E A \int_0^L \varepsilon^2 (l) \mathrm{d} l \ , \label{VEcont} \\
V_{\mathrm{B}} &=& \frac{1}{2} E I \int_0^L {\cal C}^2 (l) [ 1 + \varepsilon (l) ]^2 \mathrm{d} l \ , \label{VBcont}
\end{eqnarray}
respectively, in terms of the strain $\varepsilon (l) =  |  {\bf r}' (l) | - 1$, and curvature
${\cal C} (l) = |  {\bf r}' (l) \times  {\bf r}'' (l) | \ |{\bf r}' (l) |^{-3}$
of the filament, where the primes indicate derivative with respect to the coordinate $l$. The other parameters are the Young modulus $E$, the area $A$ and second moment of area $I$ of the transversal section of the filament.

Using a variational principle for both potential energies, the forces $ {\bf f}_{\mathrm{E}} (l)$ and ${\bf f}_{\mathrm{B}} (l)$ per unit length on an infinitesimal element of the filament can be obtained. The full expressions are rather complicate, specially for the bending term, involving up to the fourth spatial derivative of the position ${\bf r} (l)$.

We also take into account a drag force as the filament moves in an homogeneous viscous medium. The precise fluid dynamics is beyond the scope of this work, although we consider a low Reynolds number regime. Therefore, the drag force acting on each element is directly proportional to its velocity ${\bf f}_{\mathrm{vis}} (l)= - c \dot{ {\bf r}} (l)$ (the dot accounts for time derivative). The drag coefficient per unit length $c$ is proportional to the dynamical viscosity, and also depends on the geometry of the filament.

\begin{figure}
\begin{center}
\includegraphics[width=18pc]{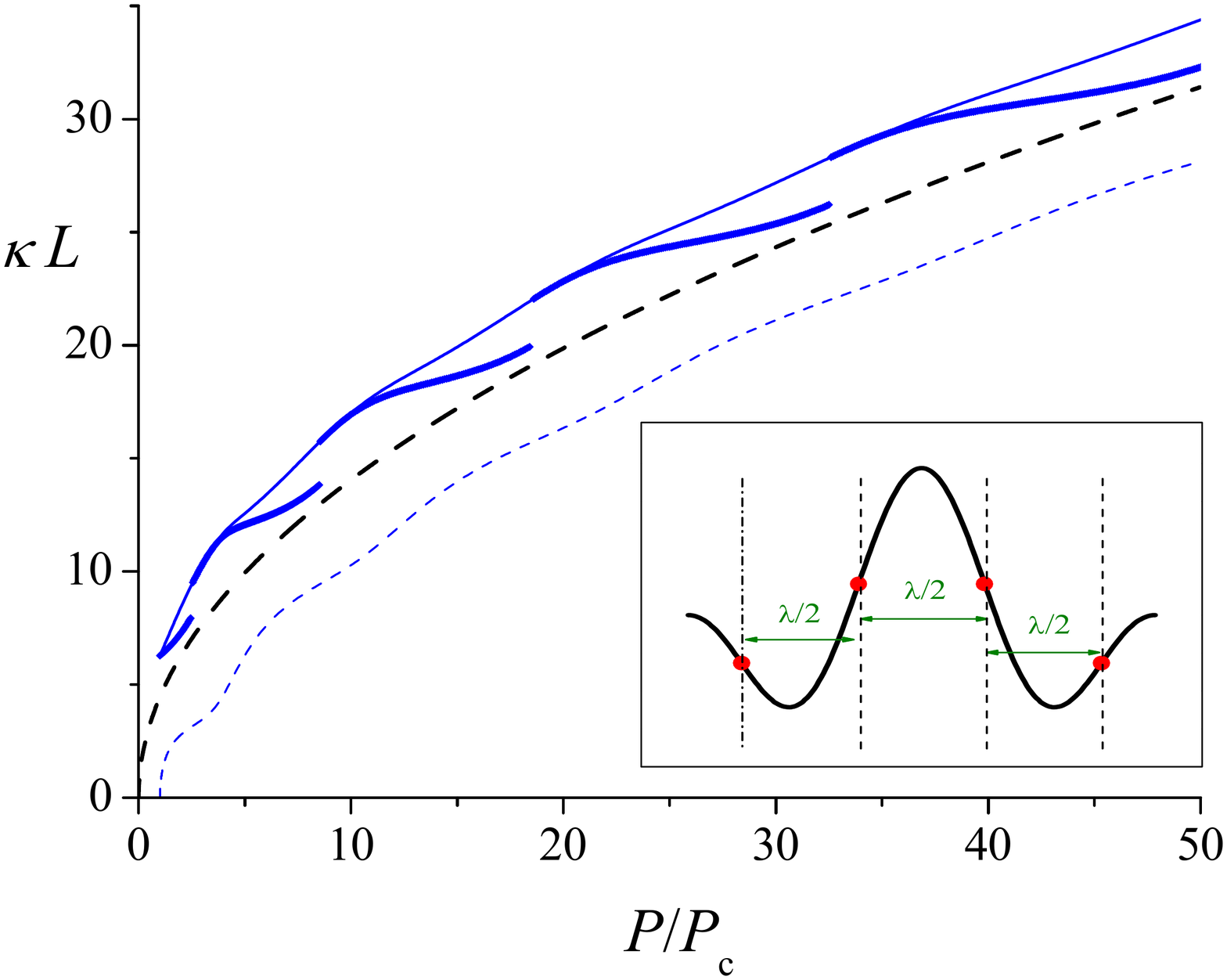}
\end{center}
\caption{\label{KappaVsForceTeo} Wavenumbers $\kappa_{-}$ and $\kappa_{+}$ (dash and continued thin lines) as a function of compressing force for $n = 1$ symmetric ($\sigma = +1$) mode. Thick line: effective wavenumber. Black dash line corresponds to mean wavenumber $\kappa = \sqrt{p/2} = \sqrt{2 \pi^2 P / P_c}$. Inset: Sketch of a normal mode and inflection points in red dots. }
\end{figure}

In an overdamped regime, the inertia term can be neglected and finally the equation of motion for an infinitesimal segment is given by

\begin{equation} \label{EqMotion}
{\bf f}_{\mathrm{E}} (l) +  {\bf f}_{\mathrm{B}} (l) +  {\bf f}_{\mathrm{vis}} (l) + {\bf f}_{\mathrm{ext}} (l) = 0 \ ,
\end{equation}
where ${\bf f}_{\mathrm{ext}}$ takes into account applied external forces. The solution is determined by the initial configuration of the filament and its boundary conditions. For small deviations from the straight shape and constant compressing load $P$, Eq.~(\ref{EqMotion}) has a general analytic solution
\begin{equation}
\delta y (l, t) = \sum_{n = 1}^{\infty} \sum_{\sigma} a^{\sigma}_n \Psi^{\sigma}_n (P, l) \exp ( \Gamma^{\sigma}_n t ) 
\end{equation}
in terms of symmetric $(\sigma = +1)$ and antisymmetric $(\sigma = -1)$ normal modes $\Psi^{\sigma}_n$, which depend parametrically on $P$ \cite{Natal2017}. For a force larger than $P_c$, there is at least one mode which exponentially grows in time ($\Gamma^{\sigma}_n > 0$). In this case the corresponding mode $\Psi^{\sigma}_n$ is a linear superposition of two sinusoidal waves with definite wavenumbers

\begin{equation}
\kappa_{\pm} = \sqrt{\frac{p \pm \sqrt{p^2 - 4 \gamma}}{2}} \label{kappa1} \ , 
\end{equation}
where $p = L^2 P / (E I)$ and $\gamma = \Gamma c L^4/(EI)$. For the clamped-clamped boundary condition the values of $\gamma$ as a function of $p$, for each mode $n$ and parity $\sigma$, can be obtained numerically finding the roots of a non-linear equation \cite{Natal2017}. For $P \gg P_c$ these values can be approximated by

\begin{equation}
\gamma^{\sigma}_n \approx \frac{p^2}{4} - \left(  n \pi \sqrt{ 2 p - 4 \pi^2 n^2 } + \sqrt{2 p} \ \epsilon^{\sigma}_n \right)^2 \ ,
\end{equation}
where $\epsilon^{\sigma}_n$ is an oscillating function of $p$, much smaller than one. Replacing this approximation in Eq.~(\ref{kappa1}), the corresponding wavenumbers are

\begin{equation}  
\kappa_{\pm} \approx \sqrt{ \frac{p}{2} \pm \sqrt{2 \pi^2 n^2 (p - 2 \pi^2 n^2)} \pm \sqrt{2 p} \epsilon^{\sigma}_n } \ .
\end{equation}
Numerically, it turns out that this approximation is very accurate even for values of $P$ near the critical value. As an example, we plot in Fig. \ref{KappaVsForceTeo} the two wavenumbers for $n = 1$ symmetric ($\sigma = +1$) mode, which have the largest growing rate $\Gamma^{\sigma}_n$.

Given one of these solutions and not knowing the compressing force, it would be difficult to determine the two characteristic wavenumbers just from its shape. This is even more difficult if there is a superposition of solutions or an arbitrary shape, as in the experimental case.

For these situations we estimate an effective wavenumber by the following criterion: for a perfect sinusoidal wave, the distance between two consecutive inflection points is trivially a half wavelength (see inset Fig.\ref{KappaVsForceTeo}). Therefore, for a particular configuration of the filament, we compute the position of $N$ inflection points, obtaining $(N-1)$ values for a half wavelength. Averaging these values, we compute an average wavelength $\bar{\lambda}$. Finally we estimate $\kappa_{\text{eff}}= 2 \pi/\bar{\lambda}$. This criterion also gives a normal standard error by the dispersion of possible different wavelengths. In Fig. \ref{KappaVsForceTeo} we plot this effective wavenumber applied to the theoretical solutions (thick line). We observe discontinuities, given by the appearance of new inflection points with the increasing force.  

This approximation may not work properly in the experimental case: observed amplitudes can be comparable to $L$, boundary conditions are more complex, and compressing force is necessarily varying in time and non homogeneous along the filament. However, these expressions give a relation between the compressing load (difficult to measure) and the observed wavenumbers in the configuration of a filament.

\section{Experiments}

The experiments were carried out inside an horizontal glass cell, open at the top, of length $L_{\text{c}} =35\pm 0.1$ cm, width $W_{\text{c}} = 15\pm 0.1$ cm, and height $H_{\text{c}} = 20\pm 0.1$ cm. The filaments used were acetate strips of rectangular section 
of width $a = 4\pm 0.1$ mm, thickness $e = 100 \pm 0.1 \mu$m, and length $L = 29\pm 0.1$ cm. Flat filaments were used to reduce possible torsions and to induce deformation in the plane perpendicular to the optical axis of the camera.

The Young's modulus of filaments was estimated by the cantilever method, by measuring the deflection of the filament at one of its ends when a 
transverse force is exerted on the free end. The value of the Young's module obtained is $2.5 \pm 0.5$ GPa.

Each end of the filament is held by two small pieces of acrylic. One of these pieces is fixed to one end of the cell, while the other end has been joined, at the opposite side, to an acrylic rod that can be moved inside the cell. The device is uniformly illuminated from the bottom of the cell using a light panel.

The used viscous medium was glycerol, whose density and viscosity at room temperature are approximately equal to 1.26 g/cm$^3$ and 1.49 Pa$\cdot$s, respectively.

\begin{figure}[htbp]
\includegraphics[width=8.0cm]{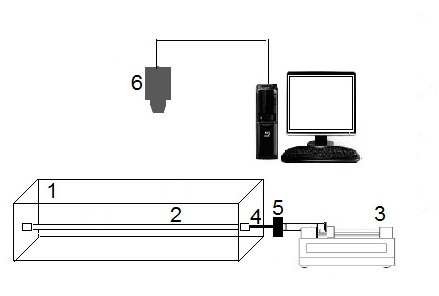}
\caption{Schematic view of the experimental set-up: 1 horizontal cell,
2 filament, 3 syringe pump, 4 acrylic rod, 5 piston, 6 camera.}
\label{fig:exp}
\end{figure}

The complete device is shown in Fig. \ref{fig:exp}.

For compression experiments the filament is placed in the holders making sure that it is completely stretched and 
without torsion. The cell is filled with glycerol until the filament 
is fully covered. Then, the filament is compressed either manually or 
by means of the piston of the syringe pump (Harvard MA 55-52226, Model 22), pushing the movable end of the filament at a constant speed. The advance of the piston stops after moving a few centimeters from the initial position to guarantee small amplitude deformations. After the piston stops, the filament relaxes.

Images are monitored and recorded during compression and relaxation with a digital camera, with spatial resolution of 0.63 mm/pixel and a frame size of 278 $\times$ 586 pixels. The images are acquired at constant time intervals (1/60 s or 1/70 s).
 
All images are processed after each experiment in order to determine the instantaneous position of the piston and the filament shape. To this purpose, we adapted the tracking routines described in \cite{Pallavicini}. Briefly, points belonging to the filament are manually selected from the initial frame and used by the code as an initial guess for the tracking. Then, the intensity profiles in the transverse direction at successive pixels along the filament extent ($x$-coordinate) are interpolated and the positions of the maximum are computed ($y$-coordinate). This procedure is repeated automatically for all the following frames in the movie. The error of the tracking was of the order of the the pixel size (0.013 cm).  The compressing velocity was measured also by image processing and included into the code, in order to define the end of the filament at each frame.

\section{Numerical simulation}

There are no analytic solutions for the hydrodynamic equation (\ref{EqMotion}) in the complex experimental situation, where one can easily control the speed of compression but not the compressing force. Nevertheless, we can check the physical model doing a finite element simulation of the filament. We divide it in $N$ equal segments of length $\Delta l = L/N$. The configuration is determined by the $(N+1)$ coordinates ${\bf r}_{n}$ of the endpoints of each segment, where $0 \leq n \leq N$. In terms of these coordinates, the elastic and bending potential energies can be written as:

\begin{eqnarray}
V_{\mathrm{E}} &=& \frac{1}{2} \frac{E A}{\Delta l} \sum_{n = 0}^{N-1} ( |{\bf r}_{n+1} - {\bf r}_{n} | - \Delta l )^2 \ , \\
V_{\mathrm{B}} &=& \frac{E I}{\Delta l} \sum_{n = 1}^{N-1} \left[ 1 - \frac{ ({\bf r}_{n+1} - {\bf r}_{n}) \cdot ({\bf r}_{n} - {\bf r}_{n-1}) }{ | {\bf r}_{n+1} - {\bf r}_{n} | \  |{\bf r}_{n} - {\bf r}_{n-1}| } \right] \ .
\end{eqnarray}
There are other possible expressions, depending on how are defined the discrete derivatives, however these expressions converge to Eqs.~(\ref{VEcont}) and (\ref{VBcont}) in the limit $\Delta l \rightarrow 0$.  Deriving the potential energies with respect to the coordinate ${\bf r}_{n}$, elastic and bending forces are obtained

\begin{eqnarray}
{\bf F}^{\mathrm{E}}_n &=& - \frac{\partial V_{\mathrm{E}}}{\partial {\bf r}_{n}} \ , \\
{\bf F}^{\mathrm{B}}_n &=& - \frac{\partial V_{\mathrm{B}}}{\partial {\bf r}_{n}} \ ,
\end{eqnarray}
which are applied to the corresponding bead $n$. The viscous force is given by ${\bf F}^{\mathrm{vis}}_n =- c \Delta l \dot{\bf r}_n$ for $1 \leq n \leq (N-1)$, which comes from the drag felt by the two adjacent semi-segments. For the end beads $n=0$ and $n = N$, the viscous force is ${\bf F}^{\mathrm{vis}}_n =- c (\Delta l / 2) \dot{\bf r}_n$, corresponding to the drag on both ending semi-segment.

Neglecting again the inertia, we arrive to $(N+1)$ coupled first-order differential equations

\begin{equation}  
\dot{\bf r}_n = \frac{\alpha_n}{c \Delta l} \left( {\bf F}^{\mathrm{E}}_n + {\bf F}^{\mathrm{B}}_n + {\bf F}^{\mathrm{ext}}_n \right) \ ,
\end{equation}
where $\alpha_n = 1$ for $1 \leq n \leq (N-1)$, and $\alpha_n = 2$ for $n=0$ and $n = N$. Given the parameters $E$, $A$, $I$, $c$, and boundary conditions, these equations are integrated numerically from a given initial configuration. For the studied velocities in the experiments, $N = 150$ segments is enough to capture the shape and observed wavenumbers.

\section{Results and Discussion}

\subsection{Estimation of the drag coefficient}

Our model depends on several parameters most of which can be estimated quite accurately from tabulated values or direct measurements. However this is not the case for the drag coefficient $c$, that have complex dependencies with the shape and inclination of the body, the Reynolds number of the flow, the roughness of the surface and the viscosity of the fluid, just to mention some.

In the experiments we observe two clearly different stages of the buckling process. The first one is related to the dynamics due to the pushing constant velocity of the piston (compression). The second one corresponds to the dynamics taken place when the piston stops and the filament shape relaxes (relaxation). The dynamical response of the filament shows to be very sensitive with the drag coefficient, playing a key role in both compression and relaxation stages. Consequently, it is mandatory to get a precise value of this parameter to achieve a more accurate description of the experimental data.

From the relaxation stage we can extract information about the characteristic time of this process, which is directly related to the drag coefficient  $c$. For this purpose we carry out the following procedure. We track the shape of the filament in the relaxation regime, for different frames of one movie at arbitrary times. Then we simulate the corresponding shapes at the same times for a given $c$ and we take the euclidean distance between the simulated and experimental contours. We repeat this procedure adjusting $c$ until the distance becomes minimum (within an uncertainty interval), obtaining an optimum $c = (91 \pm 1)$ dyn.s/cm$^2 $. We check that this value is independent of the selected times or the chosen experimental set. Then, this value is used for the simulations in the compression stage.

\subsection{Constant speed compression}

\begin{figure}[htbp]
\includegraphics[width=8.0cm]{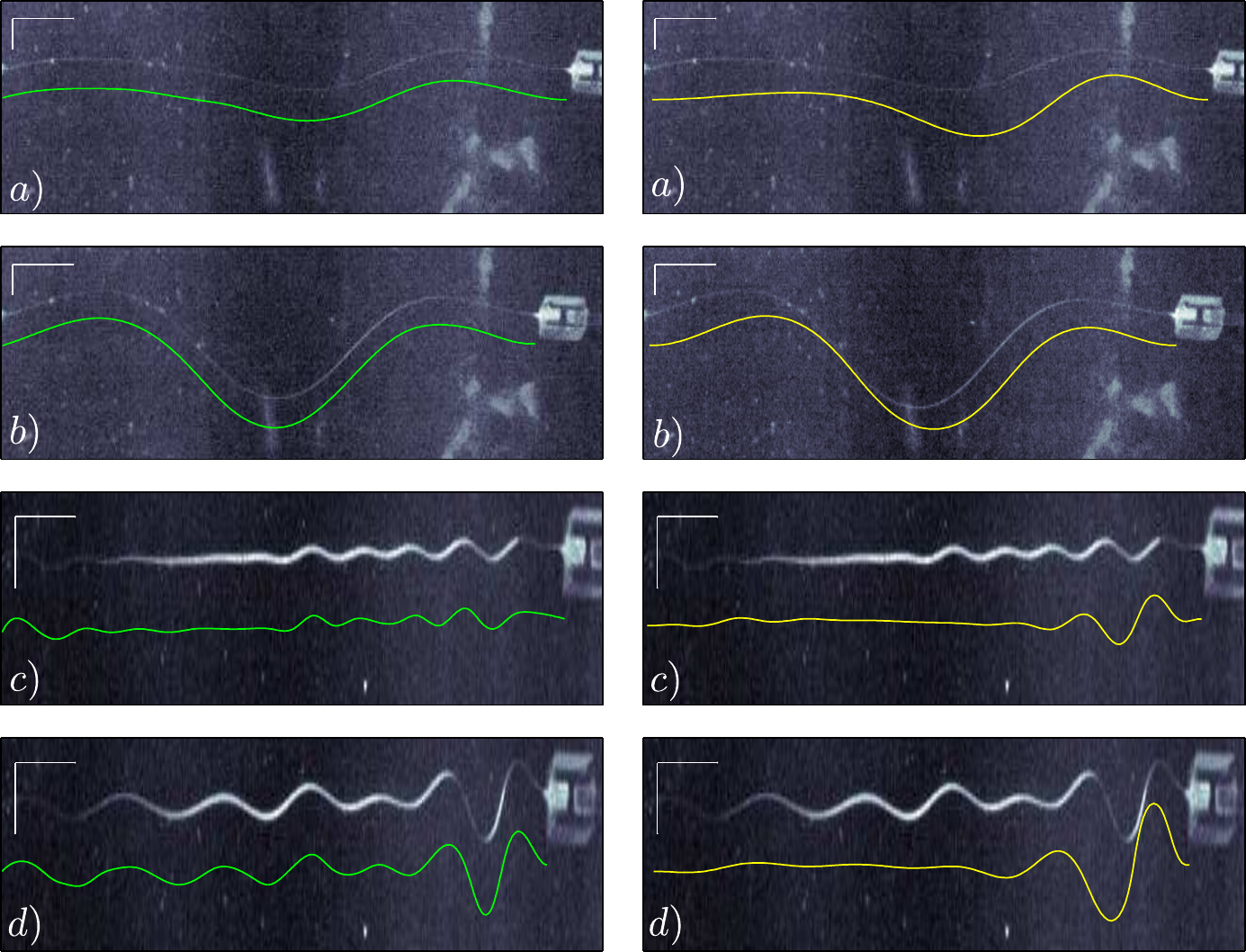}
\caption{Snapshots of two experiments. a) and b) $v_0$ = 0.062 cm/s, $t$ = 10.0 s and 33.3 s, respectively. c) and d) $v_0$ = 5 cm/s, $t$ = 0.0625 s and 0.25 s. The experimental tracking (green) and the numerical simulation (yellow) are plotted 1 cm below from the filament to facilitate visual inspection. Horizontal and vertical scale bars represent 3 cm and 1 cm respectively.}
\label{Snapshots}
\end{figure}

We analyze in detail two experiments performed at two different velocities. Initially, the filament is straight up to the tracking error. The boundary conditions are clamped-clamped (Fig.~\ref{Snapshots})

We perform numerical simulations of the buckling process with the two experimental velocities. We choose an initial configuration with an amplitude of 1 mm, which is much smaller than the length of the filament. In Fig.~\ref{Snapshots} we show experimental snapshots for two different times for each velocity, where we also plot the shapes obtained with the tracking procedure (green) and with the simulations (yellow).
Some differences can be noticed, mostly in the positions of the maxima and minima, although the amplitudes and shapes are rather similar. We do not assign these differences to a failure of the physical model, but to a sensitivity to initial configuration of the filament, as assessed numerically.  

Applying the inflection point criterion described above to the experimental and simulated shapes, we compute the effective wavenumber $\kappa $ (see Table \ref{tab_kapa}). A good agreement between simulations and experiments is found for high and low velocities.

\begin{table}[h]
    \centering
    \begin{tabular}{||c||c|c||}
    \hline
        data & $\kappa_{\text{exp}} L$&$\kappa_{\text{sim}} L$\\
        \hline
          a) & $13.8\pm 0.5$ &$12.6\pm 3.5$\\
          b) & $10.8\pm 1.6$ &$11.9\pm 1.3$\\
          c) & $62\pm18$ &$53.4\pm 7.9$\\
         d) & $46\pm15$ &$34.6\pm 5.0$\\
         \hline
    \end{tabular}
    \caption{Experimental and simulated effective wavenumber $\kappa$ for snapshots shown in Fig.\ref{Snapshots}}
    \label{tab_kapa}
\end{table}

\begin{figure}[htbp]
\includegraphics[width=8.0cm]{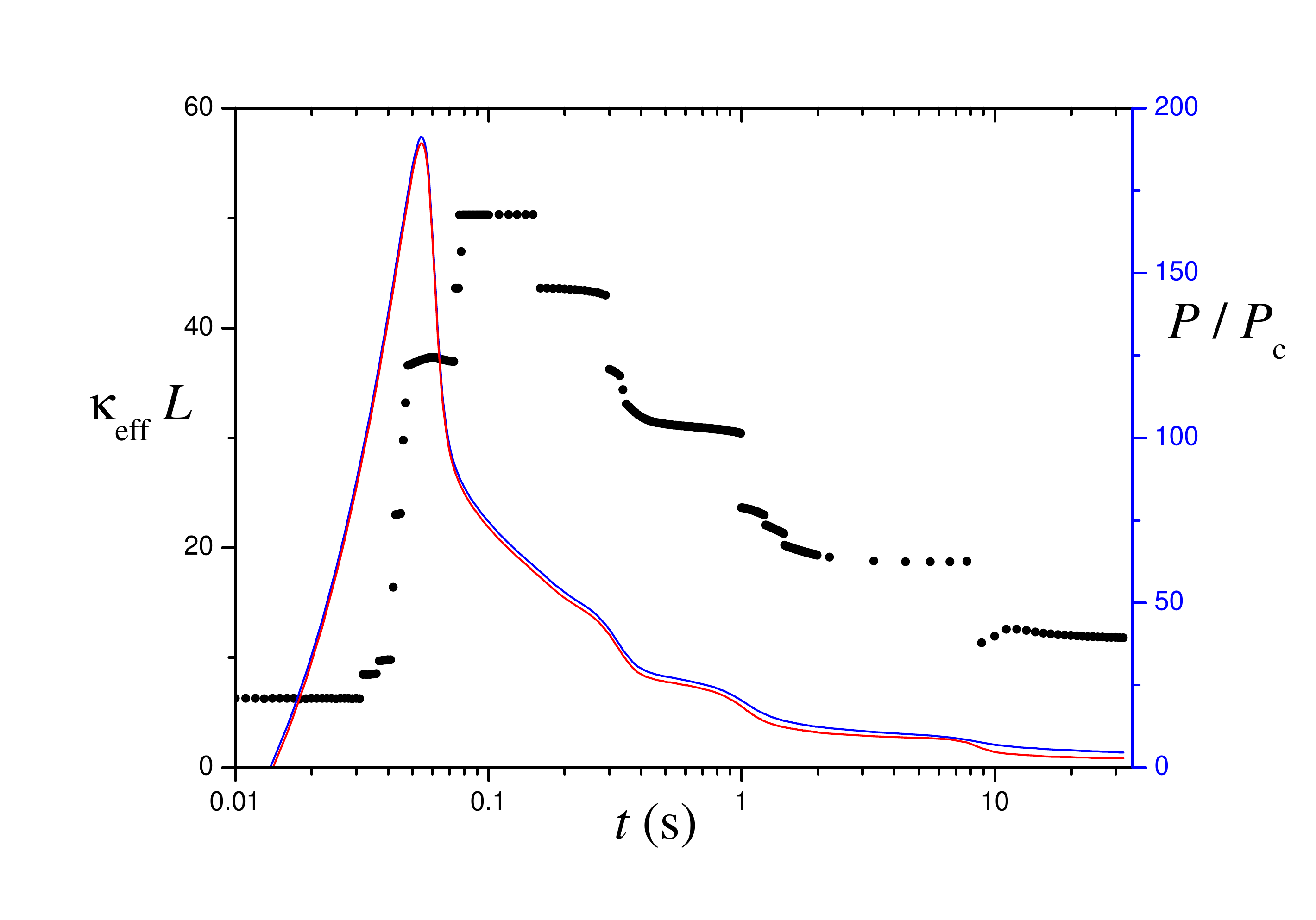}
\caption{Forces computed at the moving (red) and fixed (blue) ends of the filament, and effective wavenumber (black dots) as a function of time, for  $v_0$ = 0.062 cm/s.}
\label{KappaForceVsT}
\end{figure}

For the simulations, we also compute the longitudinal forces at the fixed and moving ends. A representative result is plotted in Fig.\ref{KappaForceVsT}.

The observed behavior is qualitatively similar for all speeds and initial shapes, provided that the initial amplitude is much smaller than $L$. For short times (notice the logarithmic scale in time), the  forces display a very rapid growth provided by an isostatic compression. During this period ($t<0.03$ s in the case shown), the filament's shape and amplitude do not vary, although the force increases several orders of magnitude. This increasing force excites high wavenumber modes that grow exponentially in time. At some point the force achieves a maximum and it decays rapidly while the amplitude increases.  As the force starts to decrease, also the wavenumbers in a step-wise manner. This non monotonic behaviour of the wavenumbers is time-shift delayed respect to the compressing force, suggesting a memory effect due to the response time scales of the different modes. In other words, there is a kind of hysteresis in the dynamical behaviour of the system.

\begin{figure}[h]
\includegraphics[width=8.0cm]{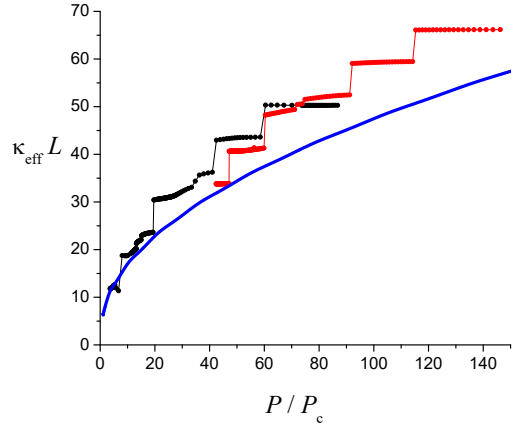}
\caption{Wavenumber as a function of average force for the two simulations of the experiments at $v_0$ = 0.062 cm/s (black dots) and $v_0$ = 5.0 cm/s (red dots). The blue curve corresponds to the theoretical value $\kappa_+$ for the symmetric mode $n=1$.}
\label{FigKappaTimeSlow}
\end{figure}

In Fig. \ref{FigKappaTimeSlow} we plot the wavenumber as a function of the force, for both theory and simulations. For the theoretical case, we plot $\kappa_{+}$ for the first symmetric mode $n=1$. For simulations we only plot data after the maximum force was attained to avoid the isostatic compression stage. Even for velocities with two orders of magnitude of difference, there is an overlap in the intermediate range of forces, indicating a similar underlying mechanism. The fact that simulations are slightly above theoretical curves is a sign of the hysteresis mentioned above.

Although our theoretical model is based on a constant compressing load, which is not the case of the experiments, it gives us a physical understanding of the relation between the wavenumbers obtained experimentally and the load, which is a great challenge to measure in experimental conditions.

\section{Conclusions}
In this paper we studied the dynamics of buckling instabilities taking place when a thin filament immersed in a very viscous fluid is compressed. 
Particularly, we explored the buckling evolution during a constant velocity compression. We determined an effective wavenumber to characterize the shape of the filament and its evolution during the compression.  By means of numerical simulation, we were able to relate the applied compressing force with the wavenumber for both high and low velocities. This is compatible with the theoretical predictions for constant loads, pointing out a similar underlying mechanism. Increasing the compressing speed results in higher forces and larger wavenumbers.  After the maximum force is reached, forces and wavenumbers evolve in a step-wise manner.

These results are interesting since they allow to estimate the values of the forces acting on a filament in a non invasive way, just measuring the wavelength from a filament's image.  This could be very useful for instance in the case of microtubules within cells, whose buckling indicates internal forces controlling important mechanical processes in the intracellular medium.

\section*{Acknowledgments}
MFC and AGM thanks PIO-CONICET grant.

\end{document}